\documentclass[prl,reprint,amsmath,amssymb,amsfonts]{revtex4-1}
\usepackage{graphicx}
\usepackage{epstopdf}
\usepackage[colorlinks=true]{hyperref}

\makeatletter
\newcommand{\customlabel}[2]{%
   \protected@write \@auxout {}{%
     \string \newlabel {#1}{{#2}{\thepage}{#2}{#1}{}}%
   }%
   \hypertarget{#1}{}%
}
\makeatother

\usepackage{mathtools}
\usepackage{dsfont}
\usepackage{bm,bbm}
\usepackage{nccmath,xfrac}
\usepackage{xcolor}
\usepackage{float}
\usepackage[final]{changes}
\usepackage{comment}

\newcommand{\n}{\bm{n}}
\newcommand{\x}{\bm{x}}
\newcommand{\z}{\bm{z}}
\newcommand{\p}{\bm{p}}
\newcommand{\dd}{\mathrm{d}}
\newcommand{\Ne}{N_{\rm e}}
\newcommand{\ve}{\varepsilon}
\newcommand{\bbar}{\bar{b}}
\newcommand{\dbar}{\bar{d}}
\newcommand{\abar}{\bar{\alpha}}
\newcommand{\gbar}{\bar{\gamma}}
\def\b{\beta}
\def\d{\delta}

\begin{document}


\title{Effective population sizes for asymmetrically regulated
  birth-death processes}


\author{Yunbei Pan$^1$}
\author{Tom Chou$^{1,2}$}
%
\affiliation{$^1$Department of Biomathematics, UCLA, Los Angeles, CA,
  90095-1766, USA}
\affiliation{$^2$Department of Mathematics, UCLA, Los Angeles, CA,
  90095-1555, USA}
\date{\today}


\date{\today}

\begin{abstract}
  \added{In multispecies birth-death processes, how population
    regulation---through suppressed replication, elevated mortality,
    or both---affects macroscopic stochastic dynamics has escaped
    detailed analysis.  Here, we show that the distribution of
    regulation mechanisms can be invisible in deterministic or
    mean-field dynamics but play a significant role in the diffusive
    evolution of population frequencies. By introducing a tunable
    regulation partitioning parameter $\alpha_i$ and projecting a
    $d$-species birth-death process onto a $(d{-}1)$-dimensional Moran
    process, we find a regulation-mechanism-dependent diffusion
    tensor.  For the simple two-species case, we derive exact fixation
    times and probabilities to show how different regulation
    mechanisms stochastically favors a more birth-regulated species,
    even under complete deterministic neutrality.
%
%
Our model also allows us to define an $\alpha$-dependent effective
population size $\Ne(\alpha)$ among neutral species, generalizing its
classical interpretation. For near-neutral populations or populations
that are heterogeneous in their regulation mechanism, we used
perturbation theory to calculate the spectral gap, identifying it with
a diversity loss timescale which can also be interpreted as setting
an effective population size. Our results are particularly applicable
to interacting subpopulations of T cells (``clones'') which are
near-neutral, are regulated through proliferation and apoptosis, and
lose diversity with time.}

\end{abstract}
\maketitle


\noindent \textbf{Introduction - }Stochastic models of the evolutionary
dynamics of multiple interacting populations typically involve
birth-death processes with nonlinear regulation effects and/or
population constraints. The simplest and most foundational of such
models are the Wright-Fisher model
\cite{fisher1923xxi,tran2013introduction,blythe2007stochastic,
  ewens2004mathematical} and its continuous-time analogue, the Moran
model \cite{moran1958random}.  A $d$-species birth-death process with
strong total population regulation can be approximated by a
$(d-1)$-dimensional Moran model in which birth and death events are
concerted to strictly fix the total population.  Note that unlike
Moran models on graphs \cite{lieberman2005evolutionary,
  zukewich2013consolidating,kaveh2015duality,kaveh2019environmental,
  svoboda2024amplifiers,brewster2025mixed}, in which the order of the
pair of birth and death events on connected nodes can lead to very
different behavior such as fixation rates, we will consider a
well-mixed population in which each birth and death are concerted.

Under a large-population ``diffusion approximation,'' the
Wright-Fisher and Moran models can be expressed as a
$(d-1)$-dimensional Fokker-Planck equation with convection and
diffusion coefficients that depend on the parameters of the underlying
birth-death process
\cite{kimura1964diffusion,traulsen2005coevolutionary,
  chalub2014frequency,black2012stochastic}. In particular, the
convection and diffusion coefficients describe the relative selection
of certain species and geometric space explored by the population
fractions.  The diffusion term also allows one to define an effective
population size that connects models that exhibit different variances
in relative populations over time
\cite{korolev2010genetic,chotibut2017population}.  For example,
effective population sizes for more complex interactions and selection
have been developed to relate discrete-time Wright-Fisher models to
continuous-time Moran models.


Here, we demonstrate that a strong total population constraint
(\textit{e.g.}, a carrying capacity) not only allows a $d$-species
birth-death process to be well approximated by a $(d-1)$-dimensional
Moran model, but that how the population regulation is implemented has
an important effect on structure of the convection and diffusion terms
in its Fokker-Planck representation.  A population may be regulated
through reduced fecundity (birth regulation), elevated mortality
(death-regulation), or some combination of both. These mechanisms are
biologically distinct: resource limitation suppressing reproduction
differs from density-dependent predation or crowding-induced
mortality. A specific motivating example is the coexisting
subpopulations of different T cells clones, each identified by
different antigen receptors. While the total population is homeostatic
\cite{surh2008homeostasis,min2018spontaneous}, under different
conditions, signaling can increase replication and/or apoptosis
\cite{hataye2006naive,schluns2003cytokine,mannering2002t,mayer2019regulation}.
For a birth-death process with linear density-dependent regulation,
the equations governing expected populations are independent of
whether regulation acts through birth, death, or any mixture
thereof. The regulation mechanism reveals itself only in the
stochastic dynamics (fluctuations) of the population(s)
\cite{chotibut2015evolutionary}.

\added{Following Constable and McKane
  \cite{constable2015models,constable2017mapping},} we analyze
$(d-1)$-dimensional Moran model approximations to the $d$-dimensional
birth-death process in the large population limit.  The diffusion
approximation leads to convection (``selection'') and diffusion
(``drift'') terms that depend not only on the individual birth rates,
death rates, and carrying capacity interaction matrix, but also on how
carrying capacity is partitioned between birth and death for each
species.  Our results reveal how selection and the ``drift'' terms,
through effective population sizes, depend on how carrying capacity
regulation \added{is implemented. By using a spectral decomposition,
  we extend the concept of effective population size to non-neutral
  systems in which species have different birth rates, death rates,
  and regulation terms. Our more general interpretation of
  effective population sizes applies to processes with separate birth
  and death events, nonlinear population regulation, and even weak
  non-neutrality, beyond its initial purpose of matching a
  discrete-generation Wright-Fisher model with the continuous time
  Moran model.}

\vspace{2mm}
\noindent \textbf{Regulated birth-death model - }We consider
$d$-species with populations $n_i$, $i=1,\ldots,d$, collected in the
state vector $\n=(n_1,\ldots,n_d)$.  The system evolves as a
continuous-time Markovian birth-death process in which each species
$i$ undergoes independent birth and death events at
\added{population-regulated per-capita rates
according to a Verhulst type model}
\begin{equation}
    \begin{aligned}
        \beta_{i}(\n) &= b_i\Big(1 - \mfrac{\alpha_{i}}{K}\sum_{j=1}^{d}\gamma_{ij}\,n_{j}\Big)^{\!+}\!n_{i}\,, \\[4pt]
        \delta_{i}(\n) &= d_{i}\Big(1 + \mfrac{(1-\alpha_{i})}{K} \mfrac{b_{i}}{d_{i}}\sum_{j=1}^{d}\gamma_{ij}\,n_{j}\Big)n_{i}\,.
    \end{aligned}
    \label{eq:rates}
\end{equation}
Here, $(\cdot)^{+}\equiv \max(\cdot, 0)$ enforces the non-negativity
of the birth rate and $b_i > 0$ and $d_i > 0$ are the intrinsic
(density-independent) per-capita birth and death rates of species
$i$. The matrix $\gamma_{ij} \sim O(1)$ represents interspecies
competition, while $K \gg 1$ sets the overall population size. Thus,
$K/\gamma_{ij}$ encodes the effective carrying capacity that species
$j$ imposes on species $i$. The parameter $\alpha_i\in [0,1]$ controls
how density-dependent regulation is partitioned between birth and
death for species~$i$: $\alpha_i = 1$ corresponds to pure
birth-regulation (density suppresses fecundity while the death rate
$\delta_i = d_i n_i$ remains density-independent), whereas $\alpha_i =
0$ corresponds to pure death-regulation (density elevates mortality
while the birth rate is unregulated). \added{The probability density
  over the population $\n$ obeys the continuous-time master equation
  given in Eq.~\ref{master0} of \ref{app:master_equation} in the
  Supplementary Material.}

The key observation is that, when $\beta_{i}(\n)>0$, the net rate,
\begin{equation}
    \beta_{i}(\n) - \delta_i(\n) = \big(b_i - d_i\big)n_i - \mfrac{b_i}{K} \sum_{j=1}^{d} \gamma_{ij}\,n_j\,n_i\, ,
    \label{eq:net_rate}
\end{equation}
is independent of $\alpha_i$. Thus, the deterministic (mean-field)
dynamics, obtained by replacing $\n$ with its expectation, are
completely blind to how regulation is distributed. In contrast, the
total event rate,
\begin{equation}
    \beta_i(\n) + \delta_i(\n) = \big(b_i + d_i\big)n_i + (1-2\alpha_i)\,\mfrac{b_i}{K}\sum_{j=1}^d\gamma_{ij}\,n_jn_i\,,
    \label{eq:total_rate}
\end{equation}
depends explicitly on $\alpha_i$. Clearly, $\alpha_{i}$ affects the
amplitude of demographic noise through the total event rate. With the
same deterministic dynamics, density regulation through death rather
than birth increases the turnover rate and produces stronger
demographic fluctuations.  \added{An explicit demonstration of how the
  regulation parameter $\alpha_{i}$ arises only in higher moments of
  the population is shown in \ref{app:master_equation}.}


\vspace{1mm}
\noindent \textbf{Diffusion approximation and projection to frequency
  simplex - } We henceforth restrict ourselves to parameter regimes in
which $b_{i} > d_{i}$, such that typical populations approach a large
carrying capacity, allowing us to apply a continuum or ``diffusion''
approximation (Eqs.~\ref{eq:FP_x} and \ref{eq:mu_sigma} in
\ref{app:Moran_projection} of the Supplementary Material.)

To isolate the effects of mean-field fitness from the microscopic
noise geometry, we further assume all intrinsic birth and death rates
$b_{i}, d_{i} \sim O(1)$ and the carrying capacity $K \gg 1$.  Then,
we consider three scenarios of parameter regimes: (i) a \textit{fully
  neutral} limit in which $b_i=b$, $d_i=d$, $\gamma_{ij}=1$, and
$\alpha_i=\alpha$, (ii) a \textit{semi-neutral} limit in which
$b_i=b$, $d_i=d$, $\gamma_{ij}=1$, with $\alpha_i\neq \alpha_j$ (for
$j\neq i$), and (iii) a \textit{quasi-neutral} case, where fitness and
regulation differences are weak: $|b_{i}-b_{j}|,\, |d_{i}-d_{j}|,
|\alpha_i-\alpha_j|, \vert \gamma_{ij} -1\vert \leq O(1/K)$.


The full $d$-dimensional birth-death process admits a natural
decomposition into a fast ``ecological'' variable $u$ (rescaled total
population) and slow ``evolutionary'' variables $p_{i}$ (species
frequencies)
\begin{equation}
    u \equiv \mfrac{1}{K}\sum_{i=1}^d n_i\,, \,\,\,
    p_i \equiv \mfrac{n_i}{\sum_{k=1}^d n_k}\,,
    \,\,\, i=1,\ldots,d-1\,,
    \label{eq:up_def}
\end{equation}
with $p_d = 1 - \sum_{i=1}^{d-1} p_i$. Consequently, the rescaled
total population $u$ is bounded above by
$u_{\max}(\p)=\min_{i}\{1/(\alpha_{i}\sum_{j}\gamma_{ij}p_{j})\}$. The
frequencies $\p = (p_1,\ldots,p_{d-1})$ live in the $(d-1)$-simplex
$\Delta^{d-1}$. The domain $\mathcal{D} = \left\{ (u, \p) \in
\mathbb{R}_{+} \times \Delta^{d-1}: 0 < u < u_{\max}(\p)\right\}$ is
thus bounded by a reflecting hypersurface $\partial\mathcal D_{\rm
  refl}=\{(u,\p):u=u_{\max}(\p)\}$. A pure death-regulation model
($\alpha_{i}\to 0$) lives in an unbounded domain. The Fokker-Planck
equation for the probability density $P(\x,t)$ (Eq.~\ref{eq:FP_x}) is
then transformed into one for the joint density $P(u,\p,t)$ given by
Eqs.~\ref{eq:app_FP_full} and \ref{eq:app_AB} in
\ref{app:Moran_projection}.

Although the extinction point $u=0$ is the true infinite-time
equilibrium, on the $O(1)$ timescale, the rescaled total population
$u$ (the ecological variable) relaxes to a local, frequency-dependent
quasi-equilibrium, $u^{\ast}(\p) = v(\p)/R(\p)$, where
$v(\p)\equiv\sum_{i=1}^d (b_i-d_i)\,p_i = \bbar(\p)-\dbar(\p)$ is the
mean net growth rate and $R(\p)\equiv\sum_{i,j=1}^d
b_i\gamma_{ij}\,p_i p_j$ is the competition-weighted birth rate. In
contrast, the frequencies $\p$ evolve on the slow timescale
$O(\sfrac{1}{K})$.  By freezing $\p$, the local conditional dynamics
of $u$ is governed by convection $A_{u} =
\sum_{i=1}^{d}(\beta_{i}-\delta_{i})/K$ and diffusion
$B_{uu}(u,\p)=\sum_{i=1}^{d}(\beta_{i}+\delta_{i})/K$. In the vicinity
of $u^{\ast}$, the linear restoring approximation $A_{u} \approx
\partial_{u}A_{u}^{\ast} \cdot (u-u^{\ast})$ balances the $O(1/K)$
demographic diffusion of magnitude $B_{uu}/(2K)$. The local dynamics
\added{of $u$} thus reduce to an Ornstein-Uhlenbeck process with
fluctuations of $O(1/\sqrt{K})$.  In the decomposition $P(u,\p,t) =
\rho(\p,t)\pi(u \vert \p)$, the conditional quasi-stationary
distribution (QSD) of $u$ converges to a sharply peaked Gaussian
\begin{equation}
  \pi(u \vert \p) \approx {\cal N} \left(u^{\ast}(\p),
  -\mfrac{B_{uu}^{\ast}(\p)}{2K\partial_{u}A_{u}^{\ast}(\p)} \right).
\end{equation}
Integrating out $u$ via the QSD yields the projected
$(d{-}1)$-dimensional Fokker--Planck equation for the marginal
frequency density $\rho(\p,t)$,
\begin{equation}
  \partial_t\rho(\p,t) + \sum_{i=1}^{d-1}\partial_{p_i}\!\big(A_i\,\rho\big)
  = \mfrac{1}{2K}\sum_{i,j=1}^{d-1}
  \partial_{p_i}\partial_{p_j}\!\big(B_{i,j}\,\rho\big)\,,
    \label{eq:moran_FP}
\end{equation}
which is a generalized ``diffusion approximation'' to a
$(d{-}1)$-dimensional Moran process with regulation-dependent
coefficients. To leading order, $A_i \equiv A_i(u^*\!,\p)$ and
$B_{i,j} \equiv B_{i,j}(u^*\!,\p)$ is evaluated at the
quasi-equilibrium population $u^*$ which is equivalent to recovering
the hard total population constraint of the standard Moran model.
%
%
Eq.~\ref{eq:moran_FP} defines the operator
$\mathcal{L}_{\alpha}^{\dagger} = - K\sum_{i=1}^{d-1} \partial_{p_i}
(A_{i}\,\cdot) +
\big(\sfrac{1}{2}\big)\sum_{i,j=1}^{d-1}\partial_{i}\partial_{j}(B_{i,j}\,
\cdot)$ and its adjoint $\mathcal{L}_{\alpha} = K\sum_{i=1}^{d-1}A_{i}
\partial_{p_i} + \big(\sfrac{1}{2}\big)\sum_{i,j=1}^{d-1} B_{i,j}
\partial_{i}\partial_{j}$.

\textit{Regulation-induced convection.} To more explicitly resolve the
convection term, first define the population averaged intrinsic birth
and death rates $\bbar(\p) \equiv \sum_{i=1}^d b_i\,p_i$ and
$\dbar(\p) \equiv \sum_{i=1}^d d_i\,p_i$, the frequency-weighted mean
regulation $\abar(\p) \equiv \sum_{i=1}^d\alpha_i\,p_i$, and the
weighted interaction coefficient $\gbar_i(\p) \equiv \sum_{j=1}^d
\gamma_{ij}\,p_j$. In the $K\gg 1$ limit, the convection coefficient
can be written as
\begin{equation}
    \begin{aligned}
      A_i(\p) &= \big[(b_i-\bbar(\p))-(d_i-\dbar(\p))\big]\,p_i \\
      \: & \quad -
        u^*(\p)\!\Big(b_i\gbar_i(\p) - \textstyle\sum_{k}b_k\gbar_k(\p)
        p_k\Big)p_i \\ \: &
       \quad + \mfrac{2b}{K}\,\big(\alpha_i -
       \abar(\p)\big)\,p_i
       + O\big(\sfrac{1}{K^2}\big)\,.
    \end{aligned}
    \label{eq:Ai}
\end{equation}
In addition to convection arising from differences in birth and death
rates, there is a regulation-induced bias of order $O(1/K)$ due to
heterogeneity in $\alpha_i$. The convection $A_{i}(\p)$ strictly
vanishes under full neutrality.

\vspace{2mm}

\textit{Regulation-mediated diffusivity.} The corresponding diffusion
tensor can be written in the form
\begin{equation}
    \begin{aligned}
        B_{i,j}(\p) = \sum_{k=1}^{d} (\mathds{1}_{ik}-p_i)(\mathds{1}_{jk}-p_j)\,p_k\,W_k(\p),
    \end{aligned}
    \label{eq:Bij_W}
\end{equation}
where $\mathds{1}_{ik}$ is the Kronecker delta and the species-specific
noise amplitude is denoted by
\begin{equation}
  W_k(\p) \equiv \frac{b_k + d_k}{u^*(\p)}
  + (1-2\alpha_k)\,b_k\gbar_k(\p)\,.
    \label{eq:Wk}
\end{equation}
Rather than a uniform scalar variance, $W_k(\p)$ explicitly summarizes
the macroscopic demographic fluctuations arising from the parameters
$b_k$, $d_k$, $\gbar_{k}(\p)$, and $\alpha_k$.


\vspace{1mm}
\noindent \textbf{Results - }\added{The evolutionary fate of the
  populations governed by Eqs.~\ref{eq:Bij_W}-\ref{eq:Wk} is shaped by
  the structure of regulation-dependent demographic noise.  The
  convection and diffusion terms in Eqs.~\ref{eq:Ai}, \ref{eq:Bij_W},
  and \ref{eq:Wk} can be further expanded in powers of $1/K$ according
  to the limits of interest, \textit{e.g.}, (i), (ii), or (iii). We
  will outline explicit results in a few select cases. For the fully
  regulated neutral case (i), the diffusion (genetic drift) is
  symmetric which allows the macroscopic fluctuations to be fully
  captured by a modified, $\alpha$-dependent scalar effective
  population size $\Ne(\alpha)$. In the semi-neutral limit (ii) with
  heterogeneous regulation $\alpha_{i}$, we find explicit results for
  $d=2$ (one-dimensional Moran model). In this case, symmetry-breaking
  regulation difference $\alpha_{1}\neq \alpha_{2}$ gives rise to both
  convection and diffusion terms that are both $O(\sfrac{1}{K})$ (an
  $O(1)$ ``Peclet number'') but explicit $\alpha_{1},
  \alpha_{2}$-dependent expressions for mean fixation times and
  fixation probabilities are found. For more general, near-neutral
  cases (iii) and higher $d\ge 3$, analytic solutions are not
  available as species-specific regulation distorts the
  multidimensional geometry of genetic drift, precluding
  identification of a $\p$-independent scalar effective population
  size.  In such cases, we propose characterizing the diversity loss
  as an overall measure of the renormalization of the diffusion
  tensor.  We define an eigenvalue based interpretation of effective
  population size through the spectral gap, which can be computed
  numerically or through perturbation theory.}

\vspace{2mm}

\textit{Fully neutral model and scalar $\Ne(\alpha)$.}  In the fully
neutral regime (i), the noise weight Eq.~\ref{eq:Wk} becomes
species-independent: $W_k \equiv W = (b+d)/u^{\ast}+(1-2\alpha)\,b$
where $u^{\ast}=(1-d/b)$. The diffusion tensor Eq.~\ref{eq:Bij_W}
therefore reduces to an isotropic form $B_{i,j} =
W\,p_i(\mathds{1}_{ij}-p_j)$. Matching the diffusion coefficient
$B_{i,j}/(2K)$ with that of the standard Wright--Fisher form
$(b+d)p_i(\mathds{1}_{ij}-p_j)/(2\Ne)$ defines a dimensionless scalar
effective population size:
\begin{equation}
  \Ne(\alpha)
  = \mfrac{\Big(\mfrac{K}{2}\Big)\Big(1-\mfrac{d^{2}}{b^{2}}\Big)}
  {1-\alpha\Big(1-\mfrac{d}{b}\Big)} = \begin{cases}
    \mfrac{K}{2}\Big(1-\mfrac{d^{2}}{b^{2}}\Big), & \!\alpha = 0\\[8pt]
    \mfrac{K}{2}\Big(\mfrac{b}{d}\Big)\Big(1-\mfrac{d^{2}}{b^{2}}\Big),
   & \!\alpha = 1. \end{cases}
%
    \label{eq:Ne}
\end{equation}
Since $b > d > 0$ and $\alpha \in [0,1]$, $\Ne$ strictly increases
with~$\alpha$, with a value that can vary by up to a factor of
$\sfrac{b}{d}$.  The limits reveal a fundamental asymmetry:
birth-regulated populations ($\alpha \to 1$) suppress replication
through density feedback without introducing additional demographic
events. This suppression results in an intrinsically lower total event
rate and weaker genetic drift compared to regulation that increases
the effective death rate. The total dynamic rate is chosen to be $b+d$
so that $N_{\rm e} = K\big((1-\sfrac{d}{b}\big)$ at
$\alpha=\sfrac{1}{2}$, when the regulation affects birth and death
rates equally.

\vspace{2mm}

\textit{Nonuniform regulation: effective selection and fixation for
  $d=2$.} For $d=2$, we can consider the more general near-neutral
birth-death process and its associated one-dimensional Moran model
defined by convection $A_1(p_1) = s_{1}(p_{1})p_{1}(1-p_{1})$ and
diffusion $B(p_1)$ expanded to $O(\sfrac{1}{K})$ and $O(1)$, respectively:

\begin{equation}
    \begin{aligned}
      s_{1}(\p) = & 
      \,(b_1-b_2)-(d_1-d_2) \\
      \: & - u^{\ast}(\p)\big[b_{1}\bar{\gamma}_{1}(\p)
        -b_{2}\bar{\gamma}_{2}(\p)\big] \\
      \: & + \mfrac{2b}{K}\,(\alpha_1-\alpha_2) + O\big(\sfrac{1}{K^2}\big)\,
%
    \end{aligned}
\end{equation}
and $B(p_1) = [(1-p_1)W_1(p_1) + p_1 W_2(p_1)]p_1(1-p_1)$, where
$p_{1}$ is the population fraction of species 1.  In this
one-dimensional model, one can derive analytic expressions for
quantities such as the fixation probability and the mean time to
fixation.  To obtain simpler expressions that focus on how
heterogeneity in regulation affects population dynamics, we restrict
ourselves to the semi-neutral case (ii) in which $b_{i}=b, d_{i}=d,
\gamma_{ij}=1$ but $\alpha_{1}, \alpha_{2}$ are arbitrary.  Then, the
selection coefficient becomes $s_{1} =
\frac{2b}{K}(\alpha_1-\alpha_2)$, and the diffusion to lowest order
simplifies to a quadratic polynomial $B(p_1) = 2b\,[b/(b-d) -
  \alpha_1(1-p_1) - \alpha_2 p_1]\,p_1(1-p_1)$.

By integrating the exact backward equation, we obtain
the closed-form fixation probability starting from an initial
frequency $p_{1}(0)=p_0$:
\begin{equation}
  q(p_0) =
  \frac{p_0(1-\alpha_2 u^{\ast})}{1 - u^{\ast}
    \big[(1-p_0)\alpha_1+p_0\alpha_2\big]}\,,\,\,\, u^{\ast}=\big(1-\sfrac{d}{b}\big),
    \label{eq:fixation}
\end{equation}
satisfying the absorbing boundary conditions $q(0)=0$ and $q(1)=1$.
This exact solution to the diffusion approximation recovers the
macroscopic neutral expectation $q(p_0)=p_0$ when $\alpha_1=\alpha_2$.

\begin{figure}[t] %
    \centering
    \includegraphics[width=\linewidth]{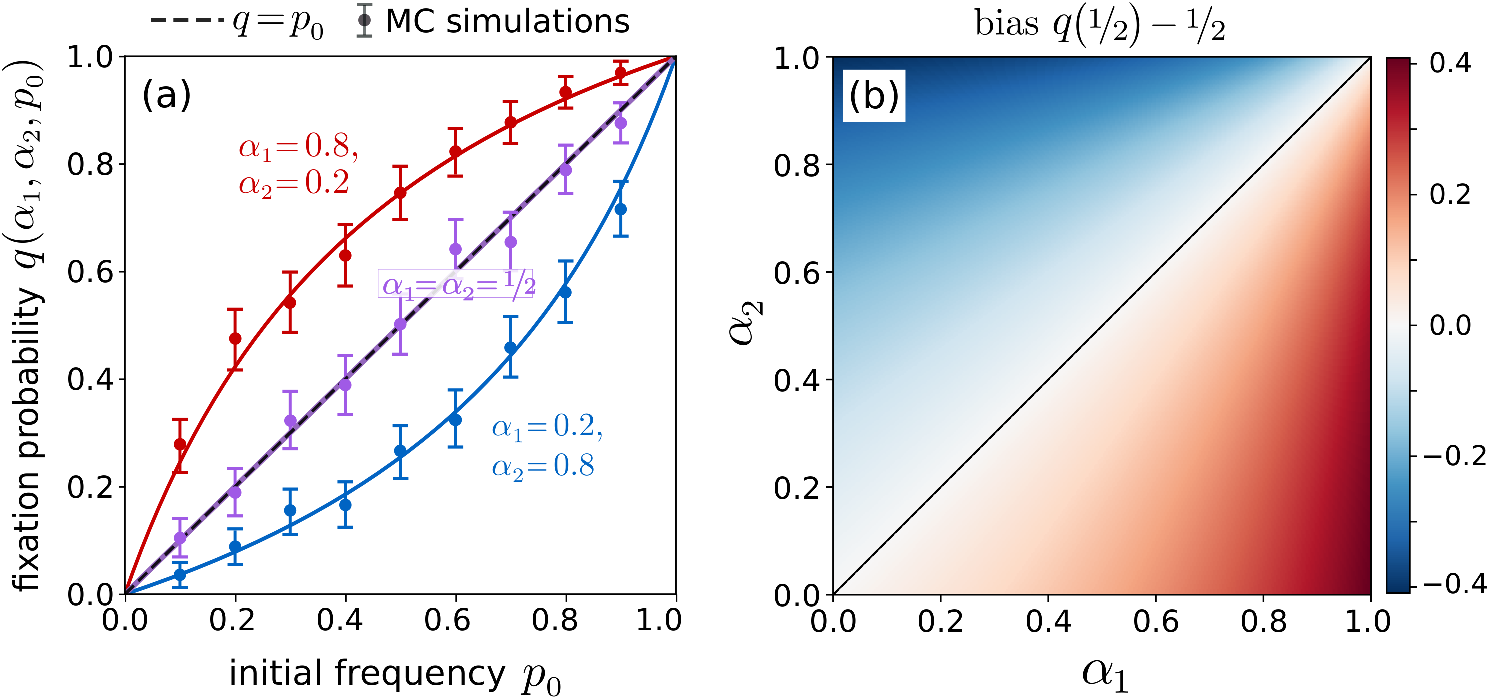}
    \caption{\textbf{Demographic noise asymmetry in $d=2$ model
        modulates extinction fates.} (a) The exact fixation
      probability $q(p_0)$ of species 1 as a function of its initial
      frequency $p_0$. Despite zero mean-field fitness differences,
      regulation asymmetry ($\alpha_1 \neq \alpha_2$) shifts the
      fixation curves away from the classical neutral baseline $q(p_0)
      = p_0$ (dashed line). Closed-form analytical predictions (solid
      lines, Eq.~\ref{eq:fixation}) match results from discrete
      Gillespie simulations, validating the diffusion approximation
      and the resulting macroscopic fixation bias. (b) Heatmap of the
      maximum fixation bias $q(1/2) - 1/2$ for species 1 across
      regulation pairings $(\alpha_1, \alpha_2)$. The species with the
      larger $\alpha$, relies relatively more on suppressed birth
      than elevated mortality, maintains a lower turnover rate, and is
      stochastically favored.}
    \label{fig:fixation}
\end{figure}

Fig.~\ref{fig:fixation}(a) shows that the species dominated by death
regulation ($\alpha_i < 1/2$) experiences stronger fluctuations and is
more likely to suffer extinction (lower fixation probability $q(p_0) <
p_0$).  Despite having no favorable selection, a birth
regulation-dominated population ($\alpha_i>1/2$) has a weaker
demographic noise amplitude, giving rise to a fixation bias $q(p_0) >
p_0$ for all $p_0 \in (0,1)$, independent of $K$. Due to the numerator
$p_0(1-p_0)$, the fixation bias ($\sim O(1)$) exactly vanishes at the
absorbing boundaries and peaks at an interior frequency, confirming
that regulation-induced selection is a global geometric property
instead of a local artifact. Fig.~\ref{fig:fixation}(b) shows a
heatmap of the maximum fixation bias $q(1/2) - 1/2$ across regulation
pairs $(\alpha_1, \alpha_2)$.
%
The expected fixation times of species 1 conditioned on fixation
$\mathds{E}[T_{1}\vert p_0]$ can also be evaluated exactly using the
conditioned backward operator, and is given in \ref{app:fixation}. The
result confirms that the stochastically favored birth-regulated
species not only has a higher probability of fixation but also
exhibits a faster fixation trajectory.

%

\vspace{2mm}

\textit{Nonuniform regulation in $d\geq 3$: Metric deformation in a
  conjugate representation.}  \added{As we have seen, in non-neutral
  cases where a frequency-independent scalar $\Ne$ cannot be uniquely
  defined, fixation times and probabilities can in principle be
  computed. Another quantity related to population size in
  high-dimensional non-neutral cases is the ``diversity loss rate''
  relative to the intrinsic diffusion rates $b_{i}+d_{i}$.  This decay
  rate is defined through the spectral gap of the Fokker-Planck
  operator.}

\added{When $d\ge 3$ and regulation $\alpha$ is heterogeneous, the
  species-specific noise weights $W_{k}$ distort the multidimensional
  geometry of diffusion. To evaluate the Fokker-Planck operator,
  either numerically or analytically, we first need to address the
  degeneracy of the diffusion tensor $B_{i,j}$ at the simplex
  boundaries $\{p_i = 0\}$, which destabilizes spatial
  discretization. Using $z_i \equiv \log(p_i/p_d), \,\,\, -\infty\leq
  z_{i}\leq +\infty$, we map these boundaries to infinity; the
  diffusion tensor takes on the form}
\begin{equation}
  \widetilde{B}_{i,j} = \mfrac{W_i}{p_i}\,\mathds{1}_{ij}
  + \mfrac{W_d}{p_d} \,
    \label{eq:Btilde}
\end{equation}
and is strictly positive definite across the entire space. The
frequencies $p_{j}(\z)$ are implicitly soft-max functions of $\z$
given in \ref{app:perturbation}. The $\p$ and $\z$ representations of
the diffusion are conjugate and preserve the common spectral
structure.

By decomposing $W_k(\p) \equiv \bar{W}(\p) + \Delta W_k(\p)$, where
$\bar{W}\equiv \sum_k p_k W_k$, the deformed metric directly inherits
the rank-one perturbation structure:
\begin{equation}
  \widetilde{B}_{i,j} = \widetilde{B}_{i,j}^{(0)} +\left(\mfrac{\Delta
    W_i}{p_i}\,\mathds{1}_{ij} + \mfrac{\Delta W_d}{p_d} \right)\,,
    \label{eq:Btilde_decomp}
\end{equation}
where $\widetilde{B}_{i,j}^{(0)} =\bar{W}\,(\mathds{1}_{ij}/p_i +
1/p_d)$ is the classical isotropic Fisher information metric recovered
in the fully neutral limit ($W_k(\p) \equiv W$). This structured
decomposition makes the regulation-induced asymmetry numerically
tractable by permitting easy application of robust spatial
discretization or adaptive spectral methods \cite{xia2021,xia2023},
enabling direct computation of the smallest positive decay rate
$\lambda_1$ for general values of $\alpha_i$, which is equivalent to
the negative of the leading nonzero eigenvalue of ${\cal L}_{\alpha}$.


\vspace{2mm} \textit{Degenerate perturbation and a spectral gap
  extension of effective population size.}  For weakly non-neutral
cases, we can more conveniently perform perturbation analysis on the
eigenvalues of the original Fokker-Planck operator ${\cal L}_{\alpha}$
using the form of ${B}_{i,j}(\p)$ given in Eq.~\ref{eq:Bij_W}.

In the fully neutral case ($W_k \equiv W$ and $\Delta W_k \equiv 0$),
the ``unperturbed'' backward operator $\mathcal{L}_{\alpha}^{(0)}$
acts on pairs of ``modes'': $\mathcal{L}_{\alpha}^{\,(0)}(p_ip_j) = -W
p_ip_j$ for any $i \neq j$.  Hence, the smallest positive decay rate
$\lambda_1^{(0)} = W$ is degenerate, spanned by the
$d(d-1)/2$-dimensional subspace of pairwise modes.  While the exact
decomposition $W_k = \bar{W} + \Delta W_k$ always holds, to provide
tractable analytic expressions, we now restrict ourselves to the
semi-neutral case with an additional constraint of weak regulation
heterogeneity $\vert \alpha_{i}-\alpha_{j}\vert <
O(\sfrac{1}{K})$. The excess term can then be treated as a
perturbation $\Delta W_k = \ve\,W_k$ with $\ve \sim 1/K$ arising from
the differences $\alpha_{i}-\alpha_{j}$. This scaling ensures that the
geometric perturbation to the demographic drift remains separated from
deterministic finite-size effects.

\ref{app:perturbation} provides details on the perturbation analysis
and shows the precise first-order splitting of the largest eigenvalue
is
\begin{equation}
  \lambda_{ij}^{(1)} = W_i + W_j
  - \left(\mfrac{1}{2}W_i + \mfrac{1}{2}W_j \right) = \mfrac{1}{2}(W_i + W_j)\,,
\end{equation}
where in this semi-neutral, weak heterogeneity case,
$(\sfrac{1}{2})W_{i}= b^{2}/(b-d) -b\alpha_{i}$, independent of $\p$.

The macroscopic diversity decay rate associated with
$\lambda_{ij}=\lambda_1^{(0)}+\lambda_{ij}^{(1)}$ splits into
$d(d-1)/2$ distinct levels. The global timescale of diversity loss
$T_{\rm div}$ is determined by the slowest decaying mode,
corresponding to the minimal spectral gap:
\begin{equation}
 T_{\rm div}^{-1} \coloneqq \lambda_1 = \min_{i \ne j} \lambda_{ij}
  = \mfrac{1}{2}\left(W_{i}^{\rm min} + W_{j}^{\rm min}\right)\,,
\end{equation}
where $W_{i}^{\rm min}$ and $W_{j}^{\rm min}$ are the two smallest
noise weights in the system. The exact closed-form expression for the
diversity loss timescale can then be used to define an
extinction-time-based spectral effective population size:
\begin{equation}
  \begin{aligned}
    \Ne^{\rm spec}(\{\alpha_{i}\}) = &
    \mfrac{2(b+d)K}{W_{i}^{\rm min} + W_{j}^{\rm min}} \\
    = &  \mfrac{\Big(\mfrac{K}{2}\Big)\Big(1-\mfrac{d^{2}}{b^{2}}\Big)}
    {1-\big(\sfrac{1}{2}\big)\big(\alpha_{i}^{\rm max}
      +\alpha_{j}^{\rm max}\big)\Big(1-\mfrac{d}{b}\Big)},
\end{aligned}
    \label{eq:Ne_spec}
\end{equation}
\added{where $\alpha_{i}^{\rm max}$ and $\alpha_{j}^{\rm max}$ are the
  two largest regulation parameters across all
  species. Eq.~\ref{eq:Ne_spec} generalizes the fully neutral
  effective population size $\Ne(\alpha)$ to take into account weakly
  heterogeneous $\alpha_{i}$ and provides an $O(1)$ correction to
  Eq.~\ref{eq:Ne}. Additional validation and connection to macroscopic
  observables are detailed in \ref{app:perturbation} of the
  Supplementary Material.}

\vspace{2mm}
\noindent \textbf{Discussion and Conclusions - }In this Letter, we
demonstrated that the microscopic mechanisms regulating multispecies
population densities are not merely ecological details that vanish
upon averaging. They affect higher correlations and shape the geometry
of the stochastic dynamics.  Classical deterministic models often
treat carrying capacity as a uniform constraint, resulting in
mean-field dynamics that cannot resolve whether a population is
regulated through suppressed fecundity or elevated mortality. By
projecting the regulated birth-death dynamics into a frequency simplex
of a diffusion approximation, we showed that heterogeneity in
regulation mechanism introduces a frequency-dependent anisotropy in
the demographic noise.

\added{To lowest order, how regulation is implemented and its
  heterogeneity across species manifests itself in the diffusion
  tensor of the population fractions. The classical reduction of
  stochastic fluctuations to a single scalar holds strictly only under
  a fully neutral model, including perfectly homogeneous regulation
  mechanisms. The regulation-dependent diffusion amplitude defines a
  generalized effective population size $\Ne(\alpha)$ which we derived
  in Eqs.~\ref{eq:Ne}.  Once species-specific regulation mechanisms
  are introduced, the isotropy of the diffusion is broken.
%
%
  In diverse ecosystems ($d \ge 3$),
  the multidimensional structure of the noise splits the macroscopic
  relaxation timescales. The loss of diversity is therefore
  governed by a spectrum of decay rates that are determined by
  the heterogeneous regulation mechanism.}

From a biological perspective, these findings show that
birth-regulated populations possess an inherent stochastic advantage
over death-regulated ones. By suppressing replication rather than
elevating mortality under crowding, birth-regulated species maintain
lower total event rates at the same carrying capacity, thereby
experiencing weaker diffusion/genetic drift. This purely noise-induced
mechanism provides a theoretical basis for understanding why certain
resilient lineages, such as memory T cells or dormant microbial
variants, might rely preferentially on division suppression rather
than apoptosis to maintain homeostasis.

The spectral framework established here provides a mathematical
foundation for the future exploration of non-neutral stochastic
processes in higher dimensions. A promising direction of investigation
is understanding how additional stochastic processes such as mutation
and environmental noise (in \textit{e.g.,} $b_{i}, d_{i}, K$) operate
under heterogeneous regulation conditions influence diversity loss
rates. Finally, our analysis assumed a linearly partitioned Verhulst
mechanism (Eq.~\ref{eq:rates}); how much of our findings persist with
other forms of population regulation should be explored.


\newpage

\bibliography{prl.bib}
\bibliographystyle{apsrev4-1}

\clearpage
\onecolumngrid

\pagenumbering{arabic}
\setcounter{page}{1}

\renewcommand{\thefigure}{S\arabic{figure}}
\setcounter{figure}{0}

\renewcommand{\theequation}{S\arabic{equation}}
\setcounter{equation}{0}

\large \noindent \textbf{Supplementary Material}

\subsection{Appendix A: Master equation and mean-field ODEs}
\customlabel{app:master_equation}{Appendix A}
%

We consider a $d$-dimensional continuous-time Markovian birth-death
process.  Generally, regulation in a birth-death process manifests
itself in population-dependent birth and death rates of each
species $i$. Assuming a Markovian birth-death process of a system in
state $n_{i},\, i=1,\ldots d$, the birth and death rates may be
written as $\b_{i}(\n)$ and $\d_{i}(\n)$, and the probability
$\mathbb{P}(\n,t)$ obeys the forward master equation
\begin{equation}
    \begin{aligned}
        \mfrac{\partial \mathbb{P}(\n,t)}{\partial t} = \sum_{i}\Big[\mathbb{A}_{i}^{-}(\beta_{i}\mathbb{P})+\mathbb{A}_{i}^{+}(\delta_{i}\mathbb{P})-(\beta_{i}+\delta_{i})\mathbb{P}\Big],
    \end{aligned}
    \label{master0}
\end{equation}
where $\mathbb{A}_{i}^{\pm}$ are the raising and lower operators on
the $i^{\rm th}$ species population $n_{i}$. This master equation can
be transformed into more tractable forms when \textit{e.g.}, the birth
and death rates lead to large populations allowing continuum
approximations. Alternatively, exact ODEs describing the evolution of
moments can be derived.

Using Eq.~\ref{eq:rates} and by applying the raising and lowering
operators of the master equation, the exact evolution of the expected
populations is governed by
\begin{equation}
  \mfrac{\dd \mathbb{E}[n_{i}]}{\dd t} =
  (b_{i}-d_{i})\mathbb{E}[n_{i}] - \mfrac{b_{i}}{K}\sum_{j} \mathbb{E}[n_{i}n_{j}]\,.
  \label{eq:mean_evolution}
\end{equation}
Neglecting correlations and approximating $\mathbb{E}[n_{i}n_{j}]
\approx \mathbb{E}[n_{i}]\mathbb{E}[n_{j}]$ yields deterministic ODEs
for $\mathbb{E}[n_{i}(t)]$. Crucially, the dependence on $\alpha_i$
strictly vanishes in the first moment, demonstrating that
deterministic (mean-field) selection is entirely blind to the
distribution of the regulation mechanism.

Dependences on the regulation mechanism $\alpha_{i}$ arise in the
second moments which evolve according to
\begin{equation}
\begin{aligned}
    \mfrac{\dd \mathbb{E}[n_{i}n_{j}]}{\dd t} = & \,\mathbb{E}\big[n_{j}\big(\b_{i}(\n)-\d_{i}(\n)\big)\big] +
    \mathbb{E}\big[n_{i}\big(\b_{j}(\n)-\d_{j}(\n)\big)\big]
    + \mathds{1}_{ij}\mathbb{E}\big[\b_{i}(\n)+\d_{i}(\n)\big] \\[6pt]
    = & \,\big(b_{i}-d_{i}+b_{j}-d_{j}\big)\mathbb{E}[n_{i}n_{j}] -\mfrac{1}{K}\sum_{\ell}\big(b_{i}\gamma_{i\ell}+b_{j}\gamma_{j\ell}\big)\mathbb{E}[n_{i}n_{j}n_{\ell}] \\[-3pt]
   \: & \hspace{2cm} + \mathds{1}_{ij}\Big[(b_{i}+d_{i})\mathbb{E}[n_{i}] + (1-2\alpha_{i})\mfrac{b_{i}}{K}\sum_{\ell}\gamma_{i\ell} \mathbb{E}[n_{i}n_{\ell}]\Big]
    \end{aligned}
    \label{eq:second_moment}
\end{equation}
where $\mathds{1}_{ij}$ is the Kronecker delta function. Note that the
last term depends on the regulation mechanism $\alpha_{i}$. When is
the choice of regulation important and how does it translate to
$(d-1)$-dimensional approximations (Moran models) in the large
$K_{i\ell}$ limit? We answer these and related questions by
investigating how stochasticity and differences in regulation
mechanisms reveal themselves under different relevant limits.

\subsection{Appendix B: Diffusion approximation and Moran limit of the birth-death process}
\customlabel{app:Moran_projection}{Appendix B}

To resolve the demographic fluctuations where $\alpha_i$ manifests, we
take the large $K$ limit and define the continuum state $\x =
\n/K$. Truncating the Kramers-Moyal expansion at the second order
yields the continuum Fokker-Planck equation:
\begin{equation}
  \partial_t P(\x,t) + \sum_{i=1}^d \partial_{x_i} [\mu_i(\x)P] = \mfrac{1}{2K} \sum_{i=1}^d \partial_{x_i}^2 [\sigma_i^2(\x)P]\,,
  \label{eq:FP_x}
\end{equation}
where the deterministic drift and  demographic variance are
\begin{equation}
    \begin{aligned}
      \mu_i(\x) \coloneqq & \mfrac{1}{K}\big(\beta_i(\x K) - \delta_i(\x K)\big) =
      (b_{i}-d_{i})x_{i} -b_{i}\sum_{j=1}^{d}\gamma_{ij}x_{i}x_{j}  \\
      \sigma_i^2(\x) \coloneqq & \mfrac{1}{K}\big(\beta_i(\x K) + \delta_i(\x K)\big)
      =(b_{i}+d_{i})x_{i} + b_{i}(1-2\alpha_{i})\sum_{j=1}^{d}\gamma_{ij}x_{i}x_{j}.
    \end{aligned}
    \label{eq:mu_sigma}   
\end{equation}

To isolate the evolutionary dynamics, \added{we transform the
  subpopulation $\x$ into a total population variable $u = \sum_{i}
  x_i$ and the frequency variables $p_i = x_i/u$ for $i=1,\ldots,d-1$
  (Eq.~\ref{eq:up_def}).} By applying It\^o's lemma, the fully coupled
  joint Fokker-Planck equation for $P(u,\p,t)$ becomes
\begin{equation}
    \begin{aligned}
      \partial_{t} P(u,\p) + \partial_u \big(A_{u}P\big)&  + \sum_{i}
      \partial_{p_{i}} \big(A_{i} P \big) \\
      \: & = \mfrac{1}{2K} \bigg[
        \partial_u^{2} \big(B_{uu}P \big) + 2 \sum_{i} \partial_{u,
          p_{i}}^{2} \big(B_{u,i} P \big) + \sum_{i,j}
        \partial_{p_{i}, p_{j}}^{2} \big(B_{i,j}P\big)\bigg],
    \end{aligned}
    \label{eq:app_FP_full}   
\end{equation}
where the transformed drift and diffusion coefficients are:
\begin{equation}
  \begin{aligned}
    A_{u}(u,\p) = & \sum_{i=1}^{d}\mu_{i}(u\p), \,\,\,
    B_{uu}(u,\p) = \sum_{i=1}^{d}\sigma_{i}^{2}(u\p)\\
    A_{i}(u,\p) = & \sum_{k=1}^{d} (\mathds{1}_{ik}-p_{i})
    \Big(\mfrac{\mu_{k}(u\p)}{u} - \mfrac{\sigma_{k}^{2}(u\p)}{Ku^{2}} \Big) \\
    B_{u,i}(u,\p) = & \mfrac{1}{u} \sum_{k=1}^{d}(\mathds{1}_{ik}-p_i)
    \sigma_{k}^{2}(u\p) \\
    B_{i,j}(u,\p) = & \mfrac{1}{u^{2}}\sum_{k=1}^{d}(\mathds{1}_{ik}-p_{i})(\mathds{1}_{jk}-p_j)
    \sigma_{k}^{2}(u\p).
 \end{aligned}
\label{eq:app_AB}
\end{equation}
After eliminating the fast variable $u$, we find the
regulation-mediated diffusion approximation to a ($d-1$)-dimensional
Moran model described by Eq.~\ref{eq:moran_FP}. Note that in the text,
we defined the species-specific noise amplitude in Eq.~\ref{eq:Wk} by
$W_k(\p) \equiv
\sigma_{k}^{2}(u^{\ast}\p)/\big(u^{\ast}(\p)\big)^{2}$.
%

%
%
%


\subsection{Appendix C: Two-species fixation and exact conditional mean first passage time}
\customlabel{app:fixation}{Appendix C}

For $d=2$, the backward equation $K A_1 q' + \frac{1}{2}B q'' = 0$
yields the exact fixation probability $q(p_0) = \int_0^{p_0}
\psi(x)\dd x / \int_0^1 \psi(x)\dd x$, $\psi(x) = \exp[-2K \int^x
  (A_1(y)/B(y))\dd y]$. IOn the semi-neutral case with heterogeneity
in only $\alpha_{i}$, the integrand $2K A_1 / B$ is a ratio of
quadratic polynomials. Its exact integration yields logarithmic
functions that, upon exponentiation into $\psi(x)$, completely
algebraically cancel, resulting in the exact rational function in
Eq.~\eqref{eq:fixation}.

To evaluate the expected time to fixation for species 1 conditioned on
fixation, we apply the Doob $h$-transform to the backward
Fokker-Planck operator $\mathcal{L}_{\alpha}$. Conditioned on reaching
the absorbing boundary at $p_0=1$, the process is governed by the
transformed operator:
\begin{equation}
  \mathcal{L}_{\alpha}^{\text{cond}}= \mfrac{1}{q(p_0)}
  \mathcal{L}\big(q(p_0)\cdot\big)
\end{equation}
Since the fixation probability satisfies the steady-state equation
$\mathcal{L}q(p_0) = 0$, expanding the product rule reduces the
conditioned operator to:
\begin{equation}
    \mathcal{L}_{\alpha}^{\text{cond}} = \mfrac{1}{2}B(p_0)\partial^2_{p_0}  + \left( A(p_0) + B(p_0)\mfrac{\partial_{p_0} q(p_0)}{q(p_0)} \right) \partial_{p_0} 
\end{equation}
The conditional mean first passage time, $\mathbb{E}[T | p_0]$,
satisfies the governing equation $\mathcal{L}^{\text{cond}}
\mathbb{E}[T | p_0] = -1$, subject to the boundary conditions that it
vanishes at the absorbing target ($\mathbb{E}[T | 1] = 0$) and remains
finite as $p_0 \downarrow 0$.

Given that $\mathcal{L}q(p_0) = 0$, we have
\begin{equation}
  \frac{1}{2} B(p_0) \left[ \partial^2_{p_0} \mathbb{E}[T | p_0]
    + \left( 2\mfrac{\partial_{p_0} q(p_0)}{q(p_0)}
    - \mfrac{\partial^2_{p_0} q(p_0)}{\partial_{p_0} q(p_0)} \right)
    \partial_{p_0} \mathbb{E}[T | p_0] \right] = -1.
\end{equation}
Applying the absorbing boundary condition $\mathbb{E}[T | 1] = 0$,
we find
\begin{equation}
  \mathbb{E}[T | p_0] = \int_{p_0}^1
  \mfrac{\partial_y q(y)}{q^{2}(y)}
  \left( \int_0^y \mfrac{2 q^{2}(x)}{B(x)
    \partial_x q(x)} \dd x \right) \dd y,
\end{equation}
which using Fubini's theorem becomes
\begin{equation}
  \mathbb{E}[T | p_0] = \int_0^{p_0} \frac{2 q(x)}{B(x) \partial_x q(x)}
  \left( \frac{1 - q(p_0)}{q(p_0)} \right) \dd x
  + \int_{p_0}^1 \frac{2 (1 - q(x))}{B(x) \partial_x q(x)} \dd x.
\end{equation}
These integrals can be evaluated to

\begin{equation}
    \begin{aligned}
      \mathbb{E}[T | p_0] = & \mfrac{1}{b \left( \mfrac{1}{u^{\ast}}
        - \alpha_2 \right)} \Bigg[ \mfrac{\mfrac{1}{u^{\ast}}
          - \alpha_1}{\alpha_1 - \alpha_2}
        \ln \left( \mfrac{\mfrac{1}{u^{\ast}} - \alpha_2}
            {\mfrac{1}{u^{\ast}} - \alpha_1} \right) -
            \mfrac{\mfrac{1}{u^{\ast}} - \alpha_1}{p_0 (\alpha_1 - \alpha_2)}
            \ln \left( \mfrac{\mfrac{1}{u^{\ast}} - \alpha_1(1-p_0)
              - \alpha_2 p_0}{\mfrac{1}{u^{\ast}} - \alpha_1} \right) \\
            \:  &\qquad\qquad\quad + \ln \left( \mfrac{\mfrac{1}{u^{\ast}} - \alpha_2}
                {\mfrac{1}{u^{\ast}} - \alpha_1(1-p_0) - \alpha_2 p_0} \right)
                + \left(1 - \mfrac{1}{p_0}\right) \ln(1 - p_0) \Bigg].
    \end{aligned}
\end{equation}

\subsection{Appendix D: Perturbation analysis of weakly deformed diffusion}
\customlabel{app:perturbation}{Appendix D}

To transform the Moran operator into a form more amenable to
perturbation analysis, we map the simplex interior
$\Delta^{d-1}_{\circ}$ diffeomorphically onto $\mathbb{R}^{d-1}$ via
the log-ratio coordinates: $z_i \equiv \log(p_i/p_d)$, $i = 1,\ldots,
d-1$.  By It\^o's calculus, the transformed diffusion tensor
$\widetilde{B}_{i,j} = \sum_{k,\ell} (\partial z_i/\partial
p_k)(\partial z_j/\partial p_{\ell})B_{k,\ell}$ analytically
simplifies, collapsing into the strictly positive-definite,
diagonal-plus-rank-one structure as given by Eq.~\ref{eq:Btilde}

\begin{equation}
  \widetilde{B}_{i,j}(\z) = \frac{W_i(\z)}{p_i(\z)}
  \mathds{1}_{ij} + \frac{W_d(\z)}{p_d(\z)},
\end{equation}
where $p_{i}(\z)$ are explicitly
\begin{equation}
p_{i}(\z) = \frac{e^{z_{i}}}{1+\sum_{j=1}^{d-1} e^{z_{j}}}, \qquad p_{d}(\z) = \frac{1}{1+\sum_{j=1}^{d-1} e^{z_{j}}}
\end{equation}
In this representation, the corresponding backward Fokker-Planck
operator, $\tilde{\mathcal{L}}_{\alpha}$, avoids finite-boundary
singularities and can be resolved using spectral methods. The
convection also accrues a geometric correction, $\widetilde{A}_i =
\sum_k (\partial z_i/\partial p_k) A_k + \frac{1}{2K}\sum_{k,l}
(\partial^2 z_i/\partial p_k \partial p_l) B_{k,l}$.

Beside permitting easier numerical solutions (for low dimensions $d$)
the $\p$ and $\z$ representations of the diffusion are conjugate and
also preserve the common spectral structure. Thus, for the
eigenvalues, we perform the perturbation analysis by evaluating the
action on the original backward operator ${\cal L}_{\alpha}$.  In the
fully neutral case ($W_k \equiv W$ and $\Delta W_k \equiv 0$), the
unperturbed backward operator $\mathcal{L}_{\alpha}^{\,(0)}$ acts on
the pairwise modes according to $\mathcal{L}_{\alpha}^{\,(0)}(p_ip_j)
= -W p_ip_j$ for any $i \neq j$.  Hence, the smallest positive decay
rate $\lambda_1^{(0)} = W$ is degenerate, spanned by the
$d(d-1)/2$-dimensional subspace of pairwise modes.

By decomposing $W_k(\p) \equiv \bar{W}(\p) + \Delta W_k(\p)$, where
$\bar{W}\equiv \sum_k p_k W_k$, the deformed metric directly inherits
the rank-one perturbation structure:
\begin{equation}
  \widetilde{B}_{i,j} = \widetilde{B}_{i,j}^{(0)} +\left(\mfrac{\Delta
    W_i}{p_i}\,\mathds{1}_{ij} + \mfrac{\Delta W_d}{p_d} \right)\,,
    \label{eq:Btilde_decomp}
\end{equation}
where $\widetilde{B}_{i,j}^{(0)} = W\,(\mathds{1}_{ij}/p_i + 1/p_d)$
is the classical isotropic Fisher information metric recovered in the
fully neutral limit ($W_k(\p) \equiv W$). This structured
decomposition makes the regulation-induced asymmetry analytically
tractable.

To be analytically explicit, we consider the weak regulation asymmetry
regime $\vert \alpha_{i}-\alpha_{j}\vert \leq O(\sfrac{1}{K})$. While
the exact decomposition $W_k = \bar{W} + \Delta W_k$ always holds, we
now restrict the magnitude of the heterogeneity such that $\Delta W_k
= \ve\,W_k$ with $\ve \sim 1/K$ arising from the differences
$\alpha_{i}-\alpha_{j}$. This scaling ensures that the geometric
perturbation to the demographic drift remains separated from
deterministic finite-size effects.  Applying the perturbed operator
$\mathcal{L}_{\alpha} = \mathcal{L}_{\alpha}^{(0)} +
\ve\,\mathcal{L}_{\alpha}^{(1)}$ to the pairwise modes yields
\begin{equation}
    \begin{aligned}
      & \mathcal{L}_{\alpha}(p_ip_j)= -p_ip_j
      \left[ W + \ve \left(W_i+W_j-\sum_{k=1}^{d}p_k W_k \right) \right].
    \end{aligned}
\end{equation}
The presence of the cubic term ($p_ip_jp_k$) breaks the closure of the
unperturbed quadratic subspace, coupling the pairwise modes to
higher-order polynomials. Standard non-degenerate theory fails.

Thus, degenerate perturbation theory is introduced by projecting the
perturbation operator $\mathcal{L}_{\alpha}^{(1)}$ onto the
$d(d-1)/2$-dimensional degenerate subspace to construct a secular
matrix.  We define the biorthogonal basis formed by the unperturbed
right (backward) eigenfunctions $\bigl|\psi_{ab}^{(0)}\bigr\rangle =
p_a p_b$ and the conjugate left (forward) eigenfunctions $\bigl\langle
\phi_{ij}^{(0)}\bigr|$, satisfying $\bigl\langle \phi_{ij}^{(0)}
\bigl| \psi_{ab}^{(0)} \bigr. \bigr\rangle = \mathds{1}_{(ij),(ab)}$.
The eigenvalue problem
$\det\bigl(\mathcal{L}_{\alpha}^{(1)}-\lambda_{1}^{(1)}\bigr)=0$
constructs the secular matrix. The first-order corrections are the
eigenvalues of the secular matrix with elements
$\mathcal{L}_{\alpha,\{(ij),(ab)\}}^{(1)} = \bigl\langle
\phi_{ij}^{(0)} \bigl| \mathcal{L}_{\alpha}^{(1)} \bigr|
\psi_{ab}^{(0)} \bigr\rangle$.

Because the system possesses strictly absorbing boundaries without
mutation, the left eigenfunction $\langle \phi_{ij}^{(0)}|$ projects
any state onto the invariant $i-j$ boundary sub-manifold. Any
polynomial containing a variable $p_k$ where $k \notin \{i,j\}$
evaluates to zero under this projection. Consequently, for any $(a,b)
\neq (i,j)$, the projection of the cubic term $\langle \phi_{ij}^{(0)}
| p_a p_b p_k \rangle$ vanishes and the secular matrix is therefore
diagonal.

To evaluate the diagonal elements
$\mathcal{L}_{\alpha,\{(ij),(ij)\}}^{(1)}$, we utilize the algebraic
closure on the $i-j$ boundary, where $p_i + p_j = 1$. The cubic
projection simplifies as $\langle \phi_{ij}^{(0)} | p_i^2 p_j \rangle
+ \langle \phi_{ij}^{(0)} | p_i p_j^2 \rangle = \langle
\phi_{ij}^{(0)} | p_i p_j (p_i + p_j) \rangle = \langle
\phi_{ij}^{(0)} | p_i p_j \rangle = 1$. By the intrinsic exchange
symmetry of the unperturbed operator, the individual projections must
be equal, yielding $\langle \phi_{ij}^{(0)} | p_i^2 p_j \rangle =
1/2$.

\begin{figure}[t] %
    \centering
    \includegraphics[width=0.8\linewidth]{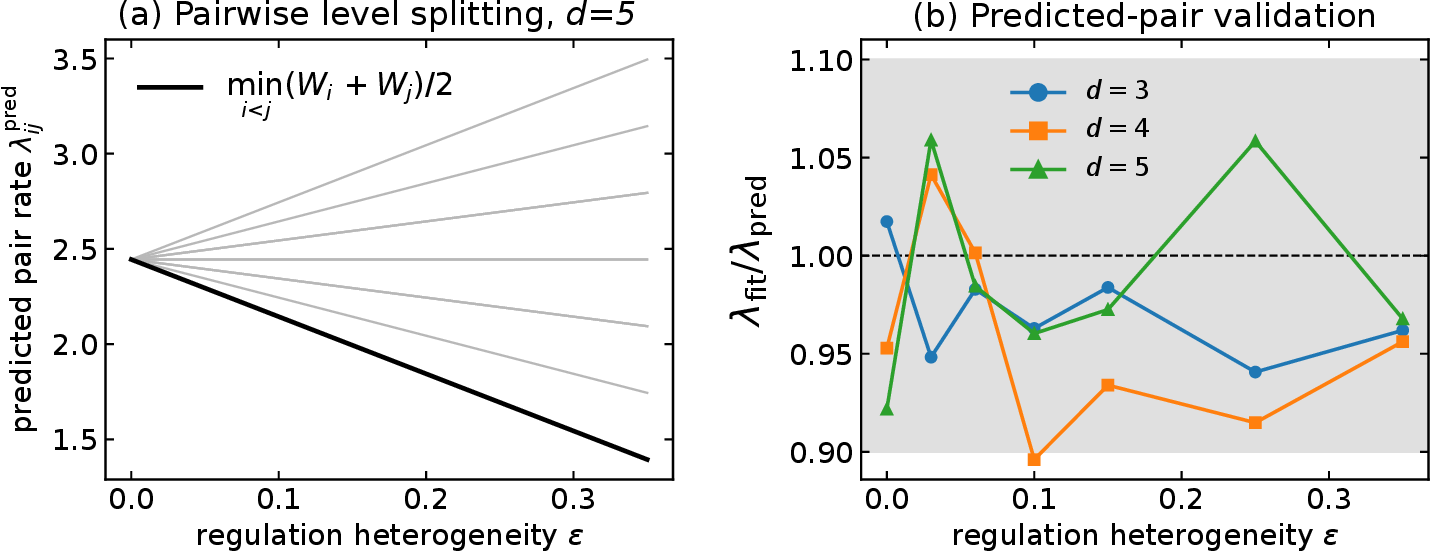}
    \caption{\textbf{Spectral gap splitting and the macroscopic
        timescale of diversity loss.} (a) Analytical prediction of the
      first-order kinetic splitting for a $d=5$ system. As regulation
      heterogeneity $\ve$ increases, the initially $d(d-1)/2$-fold
      degenerate macroscopic decay rate smoothly splits into distinct
      pairwise levels. The global timescale of diversity loss is
      governed by the minimal spectral gap (thick black line),
      corresponding to the pair with the lowest aggregate turnover
      rate.  (b) Validation of the degenerate perturbation theory
      under heterogeneous regulation ($\ve > 0$). The ratio of the
      simulated spectral gap $\lambda_{\rm fit}$ to the analytically
      predicted gap $\lambda_{\rm pred} = \frac{1}{2}(W_{\rm min} +
      W_{\rm min})$ remains near unity (mostly within the $\pm 10\%$
      gray shaded region). This confirms that the macroscopic
      bottleneck of diversity loss is precisely captured by projecting
      the metric distortion onto the pairwise subspace.}
    \label{fig:spectral}
\end{figure}

To connect this spectral gap to macroscopic observables, we monitor
the empirical pairwise heterozygosity $H(\tau) = \sum_{i < j} \langle
p_i(\tau) p_j(\tau) \rangle$ over the rescaled time $\tau =
t/K$. Because $H(\tau)$ is a linear combination of the pairwise modes,
its long-time behavior is strictly governed by the minimal spectral
gap, decaying asymptotically as $H(\tau) \sim \exp(-\lambda_1 \tau)$.
%
%
For systems with strong heterogeneous regulation ($\ve\sim O(1)$) ,
the macroscopic timescale is completely sliced apart into distinct
pairwise levels as shown in Fig.~\ref{fig:spectral}(a). The degenerate
perturbation theory accurately captures this metric distortion: the
ratio of the simulated decay rates to our analytical predictions
tightly bounds around unity (Fig.~\ref{fig:spectral}(b)). The overall
rate of diversity loss is governed by the mutually compatible species
with the lowest total event rates (the pair minimizing $W_i + W_j$),
preventing the ecosystem from rapid demographic collapse.

\end{document}